\documentclass[usegraphicx]{mn2e}

\title[The stellar mass fraction of disc galaxies]
      {From lenticulars to blue compact dwarfs: the stellar mass fraction is
       regulated by disc gravitational instability}

\author[A. B. Romeo, O. Agertz and F. Renaud]
       {Alessandro B. Romeo,$^{1}$\thanks{E-mail: romeo@chalmers.se}
        Oscar Agertz$^{2}$ and
        Florent Renaud$^{2}$\\
        $^{1}$Department of Space, Earth and Environment,
              Chalmers University of Technology,
              SE-41296 Gothenburg, Sweden\\
        $^{2}$Department of Astronomy and Theoretical Physics,
              Lund University, Box 43,
              SE-22100 Lund, Sweden}

\begin{document}

\date{Accepted 2020 October 15.
      Received 2020 October 14; in original form 2020 June 16}

\pagerange{\pageref{firstpage}--\pageref{lastpage}}

\pubyear{2021}

\maketitle

\label{firstpage}

\begin{abstract}
The stellar-to-halo mass relation (SHMR) is not only one of the main sources
of information we have on the connection between galaxies and their dark
matter haloes, but also an important indicator of the performance of galaxy
formation models.  Here we use one of the largest sample of galaxies with
both high-quality rotation curves and near-infrared surface photometry, and
perform a detailed comparative analysis of the SHMR.  Our analysis shows that
there are significant statistical differences between popular forms of the
SHMR, and illustrates the predictive power of a new physically motivated
scaling relation, which connects the stellar mass fraction
($M_{\star}/M_{\mathrm{h}}$) to the stellar specific angular momentum
($j_{\star}$) and the stellar radial velocity dispersion ($\sigma_{\star}$)
via disc gravitational instability.  Making use of such a relation, we
demonstrate (i) how challenging it is to reproduce the efficiency of galaxy
formation even for state-of-the-art cosmological hydrodynamical simulations,
and (ii) that the evolution of the stellar mass fraction is regulated by disc
gravitational instability: when $M_{\star}/M_{\mathrm{h}}$ varies,
$j_{\star}$ and $\sigma_{\star}$ also vary as predicted by our scaling
relation, thus erasing the memory of such evolution.  This implies that the
process of disc gravitational instability is intriguingly uniform across disc
galaxies of all morphological types: from lenticulars to blue compact dwarfs.
In particular, the cosmic variance of Toomre's $Q$ is 0.2 dex, a universal
value for both stars and atomic gas.
\end{abstract}


\section{INTRODUCTION}

Galaxy formation is an elaborate process involving mechanisms such as the
accretion of baryons onto dark matter haloes, galaxy mergers, cooling and
heating, gravitational instabilities, star formation, and feedback from
massive stars.  Despite this complexity, galaxies tend to be well described
by a set of scaling relations.  These are simple power laws that relate
galaxy structural parameters such as size, luminosity, rotational velocity
and angular momentum (see, e.g., Cimatti et al.\ 2020).

A fundamental aspect of galaxy formation is the connection between galaxy and
host dark matter halo properties (see, e.g., Wechsler \& Tinker 2018).  A
number of different methods are used to characterize the stellar-to-halo mass
relation (SHMR), including halo abundance matching (e.g., Vale \& Ostriker
2004; Conroy et al.\ 2006; Moster et al.\ 2010; Behroozi et al.\ 2019),
satellite kinematics (e.g., Klypin \& Prada 2009; More et al.\ 2011), weak
lensing (e.g., Mandelbaum et al.\ 2006; Leauthaud et al.\ 2012), and internal
galaxy dynamics (e.g., Read et al.\ 2017).  These techniques all
independently point to galaxy formation being a highly inefficient process,
with stars making up at most a few per cent of the halo mass in Milky
Way--like systems, and significantly less in dwarf galaxies.  Understanding
the origins of this inefficiency, and of scaling relations in general, is a
central theme in galaxy formation.

Today, several decades after the pioneering work of Ostriker \& Peebles
(1973) and the seminal papers by Efstathiou et al.\ (1982), Christodoulou et
al.\ (1995) and Mo et al.\ (1998), theoretical work is till providing fresh
insights into the link between gravitational instability and galaxy
evolution, which turn out to be useful for exploring the galaxy-halo
connection from an entirely different perspective (e.g., Athanassoula 2008;
Agertz \& Kravtsov 2016; Sellwood 2016; Okamura et al.\ 2018; Romeo \&
Mogotsi 2018; Zoldan et al.\ 2018; Valencia-Enr\'{i}quez et al.\ 2019; Romeo
2020; Zanisi et al.\ 2020).

Romeo \& Mogotsi (2018) analysed a galaxy property that is not directly
observable but is tightly constrained by gravitational instability and other
fundamental physical processes: $\langle\mathcal{Q}_{\star}\rangle$, the
mass-weighted average of the stellar $Q$ stability parameter (Toomre 1964),
corrected so as to include disc thickness effects (Romeo \& Falstad 2013).
They showed that $\langle\mathcal{Q}_{\star}\rangle$ is remarkably constant
across spiral galaxies of all morphological types (Sa--Sd),
$\langle\mathcal{Q}_{\star}\rangle\approx2\mbox{--}3$, thus proving the
existence of the self-regulation process predicted by pioneering simulation
work on spiral structure (e.g., Miller et al.\ 1970; Hohl 1971; Sellwood \&
Carlberg 1984; Carlberg \& Sellwood 1985).%
\footnote{Romeo \& Mogotsi (2018) also showed that
  $\langle\mathcal{Q}_{\star}\rangle$ is related to the Efstathiou, Lake \&
  Negroponte (1982) global stability parameter, $\epsilon_{\mathrm{m}}$, via
  the degree of rotational support, $V/\sigma$, and the velocity dispersion
  anisotropy, $\sigma_{z}/\sigma_{R}$.  These are two important effects that
  are missing from $\epsilon_{\mathrm{m}}$.}

Romeo (2020) extended the analysis carried out by Romeo \& Mogotsi (2018) to
molecular and atomic gas, and provided tighter constraints on how
self-regulated galaxy discs are.  He showed that the median value of $\langle
Q\rangle$ across the Sa--Sd sequence varies significantly not only from stars
to gas but also between the molecular and atomic gas phases, while the
$1\sigma$ scatter of $\langle Q\rangle$ is small for all the components.
Romeo (2020) further demonstrated that the low galaxy-to-galaxy variance of
$Q$ translates into a simple formula, which has important astrophysical
applications: it allows one to generate new physically motivated galaxy
scaling relations, all driven by disc gravitational instability.  An eloquent
example is provided by the atomic gas--to--stellar mass relation (see figs 4
and 5 of Romeo 2020).

In this paper, we generate one such relation and test it against two popular
forms of the SHMR.  While the present analysis builds on the approach
developed by Romeo (2020), the application presented here is new and
stretches across galaxies spanning a wide range of morphological types
(S0--BCD), stellar masses
($M_{\star}\approx10^{6.5\mbox{--}11.5}\,\mbox{M}_{\odot}$) and halo masses
($M_{\mathrm{h}}\approx10^{9\mbox{--}13}\,\mbox{M}_{\odot}$).  The rest of
the paper is organized as follows.  In Sect.\ 2, we describe the galaxy
sample, data and statistics.  In Sect.\ 3, after summarizing why such a
galaxy sample is especially appropriate for analysing the SHMR, we perform a
detailed comparative analysis of the popular $M_{\star}/M_{\mathrm{h}}$ vs
$M_{\mathrm{h}}$ and $M_{\star}/M_{\mathrm{h}}$ vs $M_{\star}$ as well as of
our scaling relation.  In Sect.\ 4, we demonstrate the importance of our
results for the simulation community, especially in the context of the
ongoing effort to reproduce the efficiency of galaxy formation in
cosmological hydrodynamical simulations.  Finally, in Sect.\ 5, we draw the
conclusions of our work and discuss its astrophysical implication.

\section{GALAXY SAMPLE, DATA AND STATISTICS}

To analyse the SHMR, we use a sample of 142 galaxies, ranging from
lenticulars to blue compact dwarfs, and spanning four and a half orders of
magnitude in $M_{\star}$ and $M_{\mathrm{h}}$.  Our galaxy sample is a
combination of the samples analysed by Posti et al.\ (2018, 2019; hereafter
P18, P19) and Iorio et al.\ (2017), which were selected from two databases:
\emph{Spitzer} Photometry and Accurate Rotation Curves (SPARC; Lelli et
al.\ 2016) and Local Irregulars That Trace Luminosity Extremes, The
H\,\textsc{i} Nearby Galaxy Survey (LITTLE THINGS; Hunter et al.\ 2012),
respectively.  Like those samples, our galaxy sample is neither statistically
complete nor volume-limited, but it is nevertheless representative of the
full population of (regularly rotating) nearby late-type galaxies (SPARC),
with an emphasis on the faint end of the luminosity function (LITTLE THINGS).
To the best of our knowledge, this is one of the largest sample of galaxies
with both high-quality rotation curves and near-infrared surface photometry
(see references above).

Our analysis requires three key quantities: $M_{\star}$, $M_{\mathrm{h}}$ and
$j_{\star}\equiv J_{\star}/M_{\star}$, the stellar specific angular momentum.
As regards the accuracy of such measurements, we distinguish four galaxy
subsamples, which we name and describe below.
\begin{enumerate}
\item `SPARC+++' contains 75 S0--BCD galaxies with accurate measurements of
  $M_{\star}$, $M_{\mathrm{h}}$ and $j_{\star}$.  The stellar and halo masses
  were measured by P19 via rotation curve decomposition, adopting a standard
  NFW halo model (Navarro et al.\ 1996) and a Bayesian approach to fit the
  observed rotation curves.  The stellar specific angular momentum was
  measured by P18 via radial integration, imposing a convergence criterion on
  the cumulative $j_{\star}(<R)$ profile and including asymmetric drift
  corrections (see, e.g., Binney \& Tremaine 2008, chap.\ 4.8.2).
\item `SPARC++' contains 35 S--BCD galaxies with accurate measurements of
  $M_{\star}$ and $M_{\mathrm{h}}$, and approximate estimates of $j_{\star}$.
  The stellar and halo masses were measured by P19, as in sample (i).  For
  these galaxies the cumulative $j_{\star}(<R)$ profile does not converge,
  hence the radial integration described in P18 would only yield a lower
  limit on $j_{\star}$.  For this reason, we estimate the stellar specific
  angular momentum adopting a commonly used approximation,
  $j_{\star}=2\,R_{\mathrm{d}}V_{\mathrm{flat}}$, where $R_{\mathrm{d}}$ is
  the exponential disc scale length and $V_{\mathrm{flat}}$ is the velocity
  along the flat part of the rotation curve (e.g., Romanowsky \& Fall 2012).
  We take $R_{\mathrm{d}}$ and $V_{\mathrm{flat}}$ from Lelli et al.\ (2016).
\item `SPARC+' contains 15 S--Im galaxies with accurate measurements of
  $j_{\star}$, and approximate estimates of $M_{\star}$ and $M_{\mathrm{h}}$.
  The stellar specific angular momentum was measured by P18, as in sample
  (i).  For these galaxies the rotation curve decomposition described in P19
  does not yield a unimodal posterior distribution for $M_{\mathrm{h}}$.
  This happens for various reasons, so Lorenzo Posti adopted various
  strategies to estimate the halo mass and kindly provided us with such
  estimates.  Nine galaxies (mostly dwarfs) have slowly rising rotation
  curves, so a cored halo model (Burkert 1995) was used to get an adequate
  fit.  Three other galaxies are nearly edge-on spirals for which the
  innermost points of the rotation curve are very uncertain, so the fit was
  repeated excluding those points.  The last three galaxies of the sample are
  spirals that have quite noisy and poorly sampled rotation curves, which
  give rise to secondary peaks in the posterior distribution of
  $M_{\mathrm{h}}$; in such a case, only the main peak of the posterior was
  considered.
\item `LITTLE THINGS' contains 17 dwarf irregular (Im) galaxies with reliable
  measurements of $M_{\star}$ and $M_{\mathrm{h}}$, and approximate estimates
  of $j_{\star}$.  The stellar and halo masses were measured by Read et
  al.\ (2017) via rotation curve decomposition.  We estimate the stellar
  specific angular momentum as in sample (ii), i.e.\ as
  $j_{\star}=2\,R_{\mathrm{d}}V_{\mathrm{flat}}$.  We take $R_{\mathrm{d}}$
  from Hunter \& Elmegreen (2006), and $V_{\mathrm{flat}}$ from Iorio et
  al.\ (2017).
\end{enumerate}

\begin{figure*}
\includegraphics[scale=1.13]{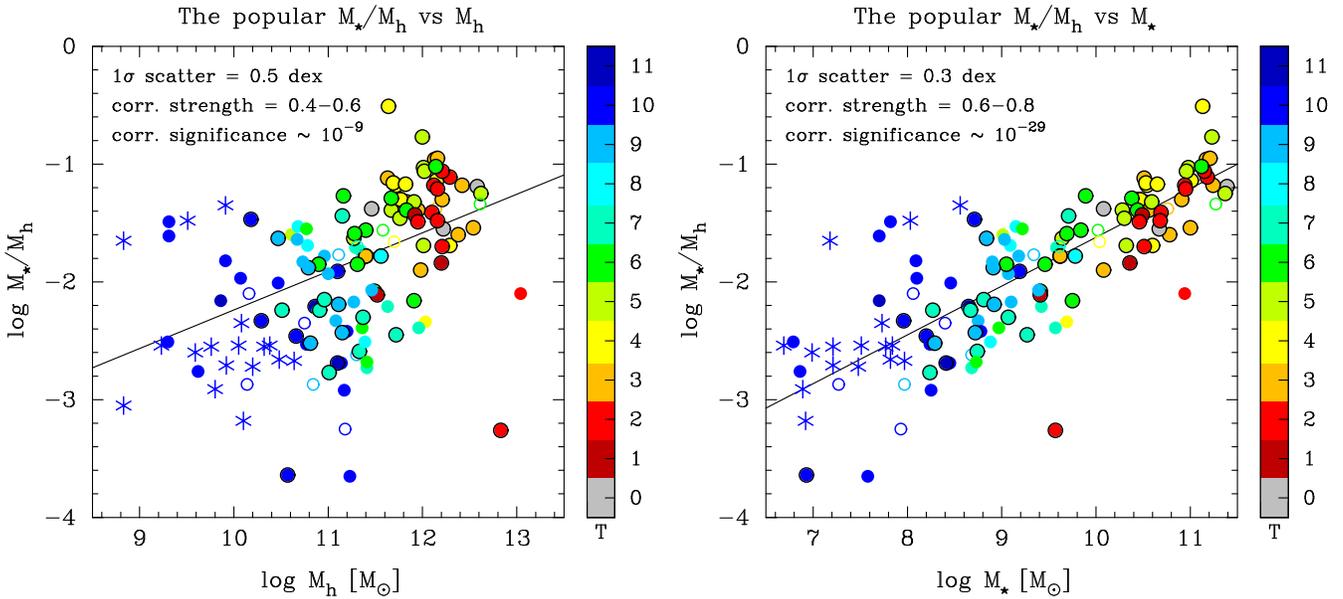}
\caption{Comparison between two popular forms of the SHMR.  Galaxies are
  colour-coded by Hubble stage, and galaxy samples are symbol-coded by the
  accuracy of the $M_{\star}$, $M_{\mathrm{h}}$ and $j_{\star}$ measurements
  (see Sect.\ 2 for more information): SPARC+++ (solid circles with black
  ouline), SPARC++ (solid circles), SPARC+ (hollow circles), LITTLE THINGS
  (asterisks).  The lines in the left and right panels,
  $\log(M_{\star}/M_{\mathrm{h}})=0.33\log M_{\mathrm{h}}-5.51$ and
  $\log(M_{\star}/M_{\mathrm{h}})=0.41\log M_{\star}-5.76$, are robust
  median-based fits to the data points.  Statistical information about the
  data is given in summary form and simplified notation (see Sect.\ 2 for
  more information).}
\end{figure*}

In addition to the key quantities specified above ($M_{\star}$,
$M_{\mathrm{h}}$ and $j_{\star}$), we need to quantify the morphological type
and the gas mass of each galaxy.  The morphological type is taken from Lelli
et al.\ (2016) for SPARC galaxies, and from Hunter et al.\ (2012) for LITTLE
THINGS galaxies.  The gas mass is computed as
$M_{\mathrm{gas}}=1.33\,M_{\mathrm{HI}}$, i.e.\ neglecting the contribution
of molecular gas (CO data are not available for most galaxies of our sample)
but including the contribution of helium to the atomic gas mass.
$M_{\mathrm{HI}}$ is taken from Lelli et al.\ (2016) for SPARC galaxies, and
from Iorio et al.\ (2017) for LITTLE THINGS galaxies.

Finally, to quantify the tightness, correlation strength and significance of
the SHMR, we present the results of several statistical measures and
associated tests.  First of all, we measure the dispersion of the data points
around the model using a robust estimator of the $1\sigma$ scatter:
$\mbox{SD}_{\mathrm{rob}}=1/0.6745\,\mbox{MAD}$, where
$\mbox{SD}_{\mathrm{rob}}$ is the robust counterpart of the standard
deviation and MAD is the median absolute deviation (see, e.g., M\"{u}ller
2000).  Values of $\mbox{SD}_{\mathrm{rob}}$ that are much less than the
dynamic range of the data mean a tight relation.  Secondly, we measure
Pearson's $r$, Spearman's $\rho$ and Kendall's $\tau$ correlation
coefficients, together with their significance levels $p_{r}$, $p_{\rho}$ and
$p_{\tau}$ (see, e.g., Press et al.\ 1992, chaps 14.5 and 14.6).  Values of
$r,\rho,\tau\approx(-)1$ and $p_{r},p_{\rho},p_{\tau}\approx0$ mean a strong
and significant (anti)correlation.  In Sect.\ 3, we will provide all such
statistical information in summary form and simplified notation.

\section{COMPARATIVE ANALYSIS OF THE SHMR}

The galaxy sample described in Sect.\ 2 is especially appropriate for
analysing the SHMR.  SPARC is one of the largest collection of galaxies with
both high-quality rotation curves and near-infrared surface photometry (Lelli
et al.\ 2016).  This allows measuring $M_{\star}$ and $M_{\mathrm{h}}$
accurately via rotation curve decomposition (P19).  The LITTLE THINGS
galaxies included in the sample also have reliable measurements of
$M_{\star}$ and $M_{\mathrm{h}}$, again derived via rotation curve
decomposition (Iorio et al.\ 2017; Read et al.\ 2017).  So, although our
sample does not contain elliptical galaxies by construction, it stretches
across all other morphological types (S0--BCD) with high data quality.  This
allows us to perform a detailed comparative analysis of the SHMR, and thus to
probe the similarities and the differences between two popular forms of the
SHMR and our scaling relation.  This section presents the results of such an
analysis.

\subsection{Two popular forms of the SHMR}

The SHMR has been massively studied adopting two alternative
parametrizations: $M_{\star}/M_{\mathrm{h}}$ vs $M_{\mathrm{h}}$ (e.g.,
Leauthaud et al.\ 2012; Behroozi et al.\ 2013; Moster et al.\ 2013;
Rodr\'{i}guez-Puebla et al.\ 2015; van Uitert et al.\ 2016; Girelli et
al.\ 2020), and/or $M_{\star}/M_{\mathrm{h}}$ vs $M_{\star}$ (e.g., Dutton et
al.\ 2010; Leauthaud et al.\ 2012; van Uitert et al.\ 2016; Lapi et
al.\ 2018; P19).  To the best of our knowledge, these two relations have not
yet been compared using data of such high quality as those described in
Sect.\ 2.  Here we perform such a comparison.

Fig.\ 1 and the statistical information shown in the two panels illustrate
that the two popular forms of the SHMR are not at all equivalent.
$M_{\star}/M_{\mathrm{h}}$ vs $M_{\mathrm{h}}$ is less constrained than
$M_{\star}/M_{\mathrm{h}}$ vs $M_{\star}$, as indicated for instance by its
larger $1\sigma$ scatter.  This is true whether we consider the whole galaxy
sample or SPARC+++, i.e.\ the galaxies with most accurate measurements of
$M_{\star}$ and $M_{\mathrm{h}}$ (as well as $j_{\star}$).  Indeed,
$M_{\star}/M_{\mathrm{h}}$ vs $M_{\mathrm{h}}$ scatters less in SPARC+++
($\mbox{SD}_{\mathrm{rob}}\approx0.4\;\mbox{dex}$) than in the whole sample
($\mbox{SD}_{\mathrm{rob}}\approx0.5\;\mbox{dex}$), but not as little as
$M_{\star}/M_{\mathrm{h}}$ vs $M_{\star}$
($\mbox{SD}_{\mathrm{rob}}\approx0.3\;\mbox{dex}$).

To understand why $M_{\star}/M_{\mathrm{h}}$ vs $M_{\mathrm{h}}$ is poorly
constrained and learn how to generate an SHMR that is tighter than
$M_{\star}/M_{\mathrm{h}}$ vs $M_{\star}$, we need to look at the SHMR from a
different perspective.  We will do this in Sects 3.2 and 3.3.

\subsection{Disc gravitational instability
            as a driver of galaxy scaling relations}

As pointed out in Sect.\ 1, Romeo (2020) demonstrated that disc gravitational
instability is a driver of galaxy scaling relations.  Here we write down the
key equation of that paper, and explain how one can use it for generating new
physically motivated scaling relations.  In Sect.\ 3.3, we will illustrate
the usefulness of such an approach in the context of the SHMR.

The origin of such relations is the low galaxy-to-galaxy variance of Toomre's
(1964) $Q$ stability parameter, which leads to Romeo's (2020) key equation:
\begin{equation}
\frac{j_{i}\hat{\sigma}_{i}}{GM_{i}}\approx1\,.
\end{equation}
Note that this is not a marginal stability condition, but a tight statistical
relation between mass ($M$), specific angular momentum ($j\equiv J/M$) and
velocity dispersion ($\hat{\sigma}$) for each baryonic component in the disc
plus bulge: atomic gas ($i=\mbox{H\,\textsc{i}}$), molecular gas
($i=\mbox{H}_{2}$) and stars ($i=\star$).  More precisely, $\hat{\sigma}_{i}$
is the radial velocity dispersion of component $i$, $\sigma_{i}$, properly
averaged and rescaled.  This quantity can be evaluated using two alternative
equations, depending on whether there are reliable $\sigma_{i}$ measurements
available or not.  Unfortunately, such measurements are highly non-trivial
(e.g., Ianjamasimanana et al.\ 2017; Marchuk \& Sotnikova 2017), hence very
sparse (e.g., Romeo \& Mogotsi 2017; Mogotsi \& Romeo 2019).  Therefore, if
one wants to analyse a large galaxy sample, then the appropriate equation to
use is
\begin{equation}
\hat{\sigma}_{i}=
\left\{
\begin{array}{rll}
 11\;\mbox{km\,s}^{-1}
&
& \mbox{if\ }i=\mbox{H\,\textsc{i}}\,, \\
  8\;\mbox{km\,s}^{-1}
&
& \mbox{if\ }i=\mbox{H}_{2}\,, \\
130\;\mbox{km\,s}^{-1}
& \!\!\!\!\times\,\,(M_{\star}/10^{10.6}\,\mbox{M}_{\odot})^{0.5}
& \mbox{if\ }i=\star\,.
\end{array}
\right.
\end{equation}
Note that these are not observationally motivated values of the gas and
stellar velocity dispersions, but rigorously derived values of the velocity
dispersion--based quantity $\hat{\sigma}_{i}$.

To make good use of Eq.\ (1), one needs to understand three key points:
\begin{itemize}
\item Eq.\ (1) means that $j_{i}\hat{\sigma}_{i}/G$ is an accurate predictor
  of $M_{i}$.  In contrast to other mass estimators,
  $j_{i}\hat{\sigma}_{i}/G$ is entirely based on the dynamics of galaxy
  discs, thus it does not rely on any assumption about the
  CO-to-$\mathrm{H}_{2}$ conversion factor ($X_{\mathrm{CO}}$) or the stellar
  mass-to-light ratio ($\Upsilon_{\star}$).
\item Eq.\ (1) can be turned into an equally accurate mass fraction
  predictor.  Simply multiply the left- and right-hand sides by
  $M_{i}/M_{\mathrm{ref}}$, where $M_{\mathrm{ref}}$ is any well-defined
  reference mass.  By construction, the resulting scaling relation,
  $M_{i}/M_{\mathrm{ref}}$ vs $GM_{\mathrm{ref}}/j_{i}\hat{\sigma}_{i}$, will
  have the same (logarithmic) scatter as Eq.\ (1), which itself is identical
  to the galaxy-to-galaxy scatter of $Q_{i}$.  Clearly, the tightness of the
  resulting scaling relation will depend not only on its scatter, but also on
  the dynamic range of $M_{i}/M_{\mathrm{ref}}$: the wider the range, the
  tighter the relation.
\item Eq.\ (1) applies to spiral galaxies of all morphological types
  (Sa--Sd).  The tests carried out by Romeo (2020) suggest that Eq.\ (1) has
  a wider range of applicability (Sa--dIrr), but so far conclusive evidence
  has only been found for the atomic gas relation (Eq.\ 1 for
  $i=\mbox{H\,\textsc{i}}$).
\end{itemize}

\subsection{Our scaling relation, and
            the roles of specific angular momentum and velocity dispersion}

Now that we know how to make good use of Eq.\ (1), let us apply such a simple
formula in practice.  Choosing the mass of the dark matter halo as the
reference mass, we get the following predictor for the stellar mass fraction:
\begin{equation}
\frac{M_{\star}}{M_{\mathrm{h}}}\approx
\frac{j_{\star}\hat{\sigma}_{\star}}{GM_{\mathrm{h}}}\,.
\end{equation}
Before testing Eq.\ (3), let us clarify two important points:
\begin{itemize}
\item Since $M_{\mathrm{h}}$ plays the role of a reference mass, Eq.\ (3)
  contains the same information as the original Eq.\ (1) [$i=\star$].  This
  apparently pure stellar equation encloses two layers of information about
  the dark matter halo.  The first layer concerns $j_{\star}$.  By
  definition, $j_{\star}$ is the stellar mass--weighted average of
  $RV_{\mathrm{rot}}(R)$, the orbital specific angular momentum.  Hence
  $j_{\star}$ depends not only on the mass distribution of the stellar disc,
  but also on the gravitational potential of the halo.  The second layer of
  information is deeper: it concerns the origin of Eq.\ (1) [$i=\star$],
  which is the remarkably flat radial distribution of $Q_{\star}$ (see
  fig.\ 1 of Romeo 2020).  The flatness of $Q_{\star}(R)$ indicates that
  stellar discs self-regulate their mass surface density and radial velocity
  dispersion so as to lie at the edge of gravitational instability, a state
  that is dynamically influenced by the gravitational potential of the host
  dark matter haloes (see again Romeo 2020).
\item Since Eq.\ (3) contains only implicit information about the dark matter
  halo, $M_{\star}/M_{\mathrm{h}}$ vs
  $GM_{\mathrm{h}}/j_{\star}\hat{\sigma}_{\star}$ is perhaps not a proper
  SHMR, but it is nevertheless a physically motivated scaling relation that
  can provide new insights into the galaxy-halo connection.  In fact,
  $M_{\star}/M_{\mathrm{h}}$ vs
  $GM_{\mathrm{h}}/j_{\star}\hat{\sigma}_{\star}$ can be directly compared
  with the popular $M_{\star}/M_{\mathrm{h}}$ vs $M_{\mathrm{h}}$, which is
  important not only for understanding why $M_{\star}/M_{\mathrm{h}}$ vs
  $M_{\mathrm{h}}$ is poorly constrained but also for exploring the roles
  that $j_{\star}$ and $\hat{\sigma}_{\star}$ play in such a scenario.
\end{itemize}

\begin{figure*}
\includegraphics[scale=1.13]{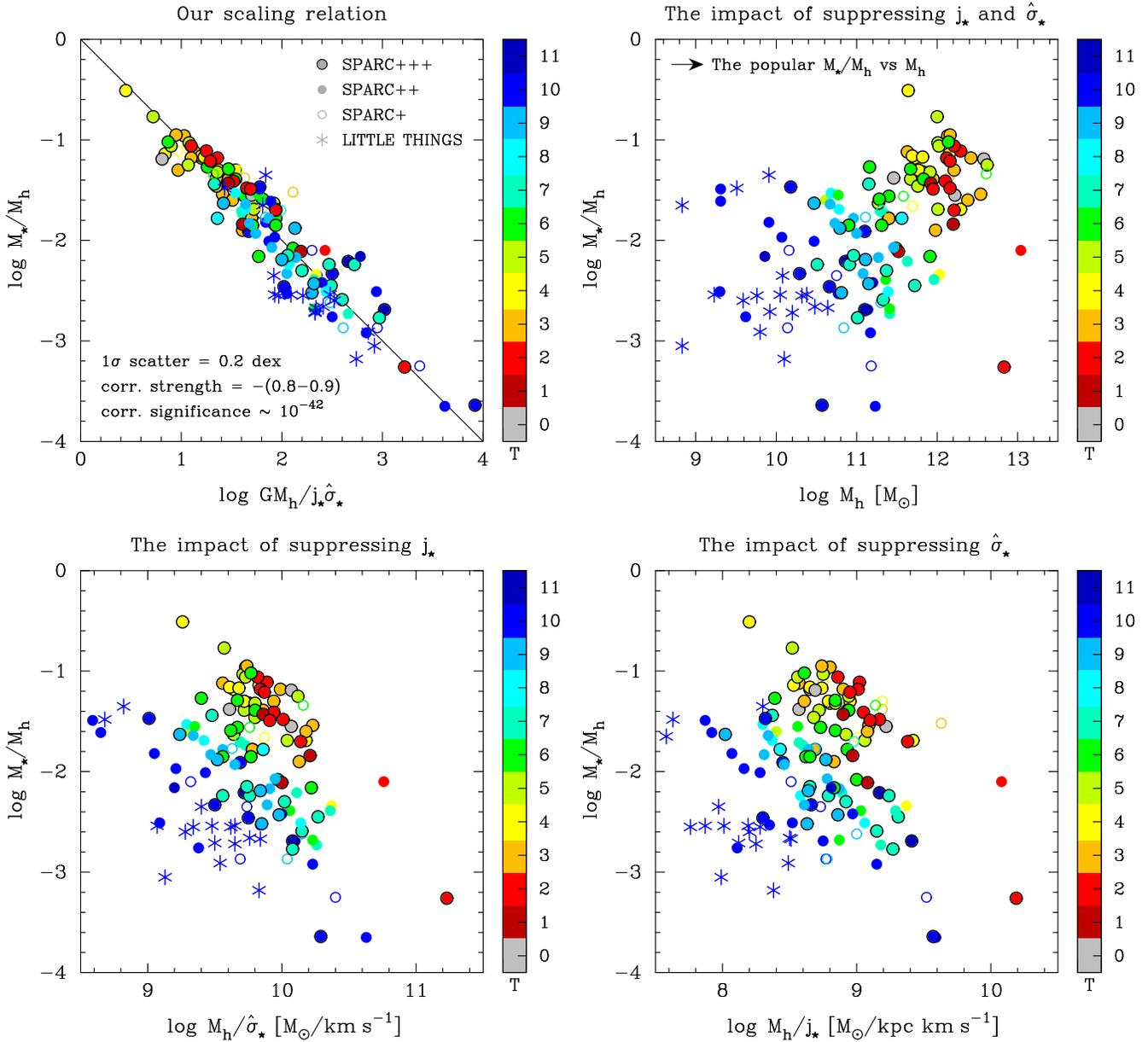}
\caption{Our scaling relation, $M_{\star}/M_{\mathrm{h}}$ vs
  $GM_{\mathrm{h}}/j_{\star}\hat{\sigma}_{\star}$, and the impact of
  suppressing $j_{\star}$ and/or $\hat{\sigma}_{\star}$.  Galaxies are
  colour-coded by Hubble stage, and galaxy samples are denoted as in
  Sect.\ 2.  In the top-left panel, the diagonal line is the prediction based
  on disc gravitational instability, and statistical information about the
  data is given in summary form and simplified notation (see Sect.\ 2 for
  more information).}
\end{figure*}

Fig.\ 2 shows our scaling relation, $M_{\star}/M_{\mathrm{h}}$ vs
$GM_{\mathrm{h}}/j_{\star}\hat{\sigma}_{\star}$, and the impact of
suppressing $j_{\star}$ and/or $\hat{\sigma}_{\star}$.  Let us first focus on
the top-left panel.  A comparison with Fig.\ 1 illustrates that our scaling
relation is more tightly constrained than the popular
$M_{\star}/M_{\mathrm{h}}$ vs $M_{\mathrm{h}}$ and $M_{\star}/M_{\mathrm{h}}$
vs $M_{\star}$.  Indeed, it is striking that a simple formula like Eq.\ (3),
entirely based on a theoretical approach, succeeds in predicting the stellar
mass fraction across four orders of magnitude in $M_{\star}/M_{\mathrm{h}}$,
to within 0.2 dex, and without any free parameter or fine-tuning.  These
facts speak clearly.  In particular, the small scatter of our scaling
relation tells us that the gravitational instability properties of galaxy
discs are, on average, remarkably uniform across the sequence S0--BCD.  This
confirms and extends a result originally found by Romeo (2020) in the context
of the atomic gas--to--stellar mass relation.  The facts pointed out above
are true whether we consider the whole galaxy sample or SPARC+++.  While
$M_{\star}/M_{\mathrm{h}}$ vs $M_{\mathrm{h}}$ and $M_{\star}/M_{\mathrm{h}}$
vs $M_{\star}$ scatter less in SPARC+++ than in the whole sample (see
Sect.\ 3.1), our scaling relation is insensitive to the accuracy of the
measurements and, whichever (sub)sample is considered, it outperforms the two
popular relations in terms of tightness, correlation strength and
significance.

There is one more point that we want to understand about our scaling
relation.  Since it connects three basic quantities (mass, specific angular
momentum and velocity dispersion), and since mass is generally regarded as
the most fundamental one, we ask ourselves: which of the other two quantities
plays a more important role in our prediction?  Fig.\ 2 answers this
question.  Suppressing $j_{\star}$ and/or $\hat{\sigma}_{\star}$ clearly has
a strong impact on the result.  In particular, suppressing both such
quantities results in the popular $M_{\star}/M_{\mathrm{h}}$ vs
$M_{\mathrm{h}}$ (see top-right panel), which is as poorly constrained as the
relations shown in the bottom panels.  Thus specific angular momentum and
velocity dispersion play an equally important role in our prediction, and it
is their interplay that constrains the SHMR so tightly in such a scenario.

\subsection{Residuals of the three SHMRs}

To fully characterize the three SHMRs, we have analysed their residuals,
$\Delta\log(M_{\star}/M_{\mathrm{h}})$, i.e.\ the deviations of the
$\log(M_{\star}/M_{\mathrm{h}})$ measurements from the underlying linear
model (a line in a log-log plot).  Remember that the model shown in the
top-left panel of Fig.\ 2 is the prediction based on disc gravitational
instability, while the models shown in the left and right panels of Fig.\ 1
are robust median-based fits to the data points (see figure caption for more
information).  Fits based on robust statistics like the median are especially
useful when the data are few or contain a significant fraction of outliers,
or even when the data deviate significantly from a normal distribution (see,
e.g., Rousseeuw 1991; Press et al.\ 1992, chap.\ 15.7).

\begin{figure*}
\includegraphics[scale=.92]{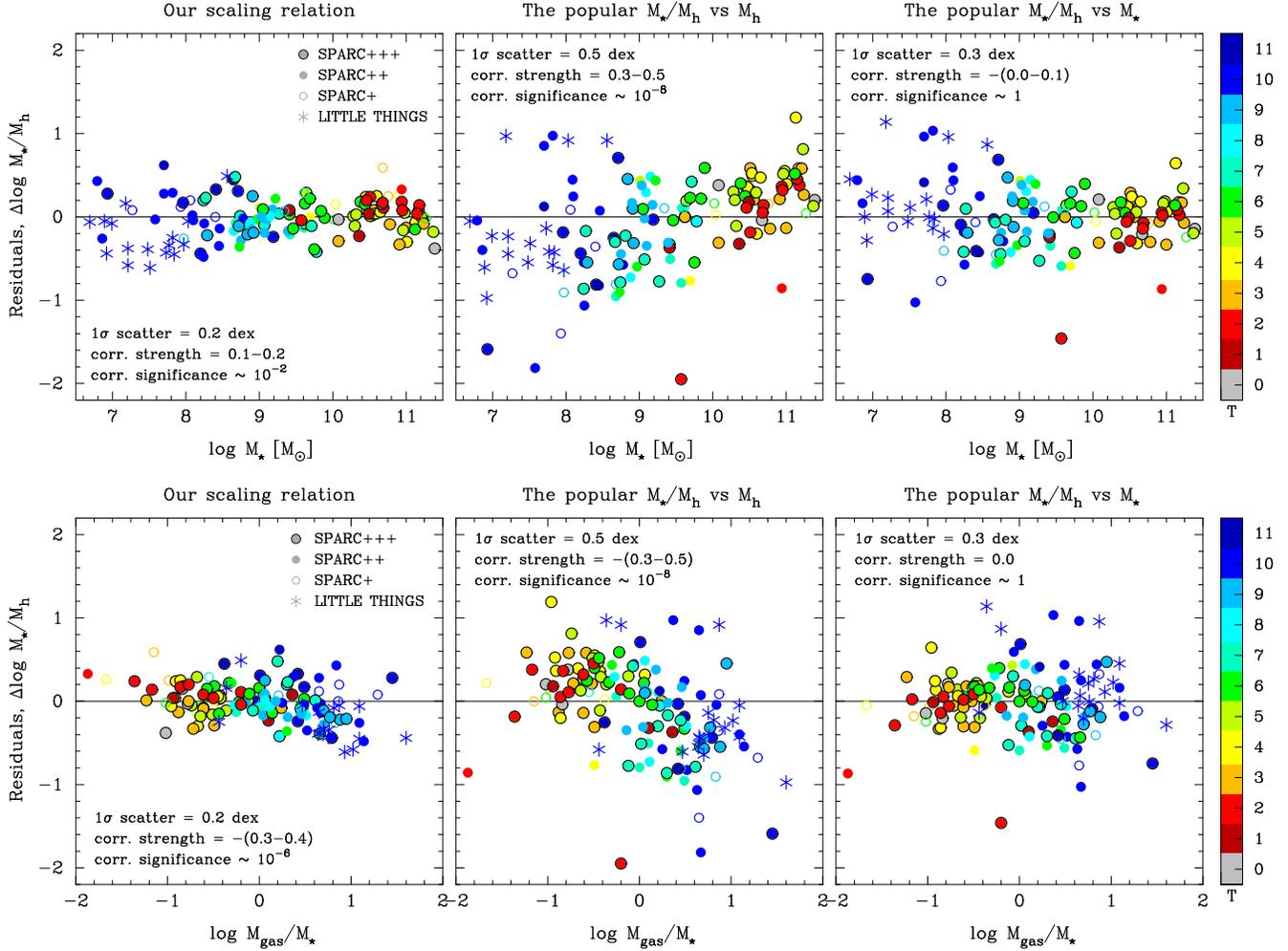}
\caption{Residuals of the three SHMRs versus $M_{\star}$ (top panels) and
  $M_{\mathrm{gas}}/M_{\star}$ (bottom panels).  Galaxies are colour-coded by
  Hubble stage, and galaxy samples are denoted as in Sect.\ 2.  Statistical
  information about the data is given in summary form and simplified notation
  (see Sect.\ 2 for more information).}
\end{figure*}

The results of our analysis are illustrated in Fig.\ 3, both as a function of
$M_{\star}$ (top panels) and as a function of $M_{\mathrm{gas}}/M_{\star}$
(bottom panels).  Our scaling relation shows a residual correlation with
$M_{\mathrm{gas}}/M_{\star}$, but this is a second-order effect that depends
on the accuracy of the measurements, and which vanishes in SPARC+++.  The two
popular relations disclose an interesting difference:
$M_{\star}/M_{\mathrm{h}}$ vs $M_{\star}$ shows no residual correlation at
all, while $M_{\star}/M_{\mathrm{h}}$ vs $M_{\mathrm{h}}$ shows residual
correlations with both $M_{\star}$ and $M_{\mathrm{gas}}/M_{\star}$.  This is
true whether we consider the whole galaxy sample or SPARC+++, and highlights
two important facts.  First, $M_{\star}$ is a more reliable estimator of
$M_{\star}/M_{\mathrm{h}}$ than $M_{\mathrm{h}}$: it is unbiased, and it also
has less scatter.  Second, the large scatter of $M_{\star}/M_{\mathrm{h}}$ vs
$M_{\mathrm{h}}$ hides a weak residual trend with stellar mass and reverse
trend with gas mass fraction, i.e.\ massive gas-poor galaxies (lenticulars
and early-type spirals) tend to have a higher stellar-to-halo mass ratio than
expected, and vice versa for low-mass gas-rich galaxies (late-type spirals
and dwarfs).

\section{OUR DIAGNOSTICS AS A CRITICAL TEST
         FOR SIMULATIONS OF GALAXY FORMATION AND EVOLUTION}

The SHMR is not only one of the main sources of information we have on the
galaxy-halo connection, but also an important indicator of the performance of
galaxy formation models (e.g., Dutton et al.\ 2011; Read et al.\ 2017; Forbes
et al.\ 2019; Agertz et al.\ 2020; Marasco et al.\ 2020, hereafter M20;
Rodr\'{i}guez-Puebla et al.\ 2020; Zanisi et al.\ 2020).  Here we join this
effort and demonstrate the importance of our results for the simulation
community.

\begin{figure*}
\includegraphics[scale=1.22]{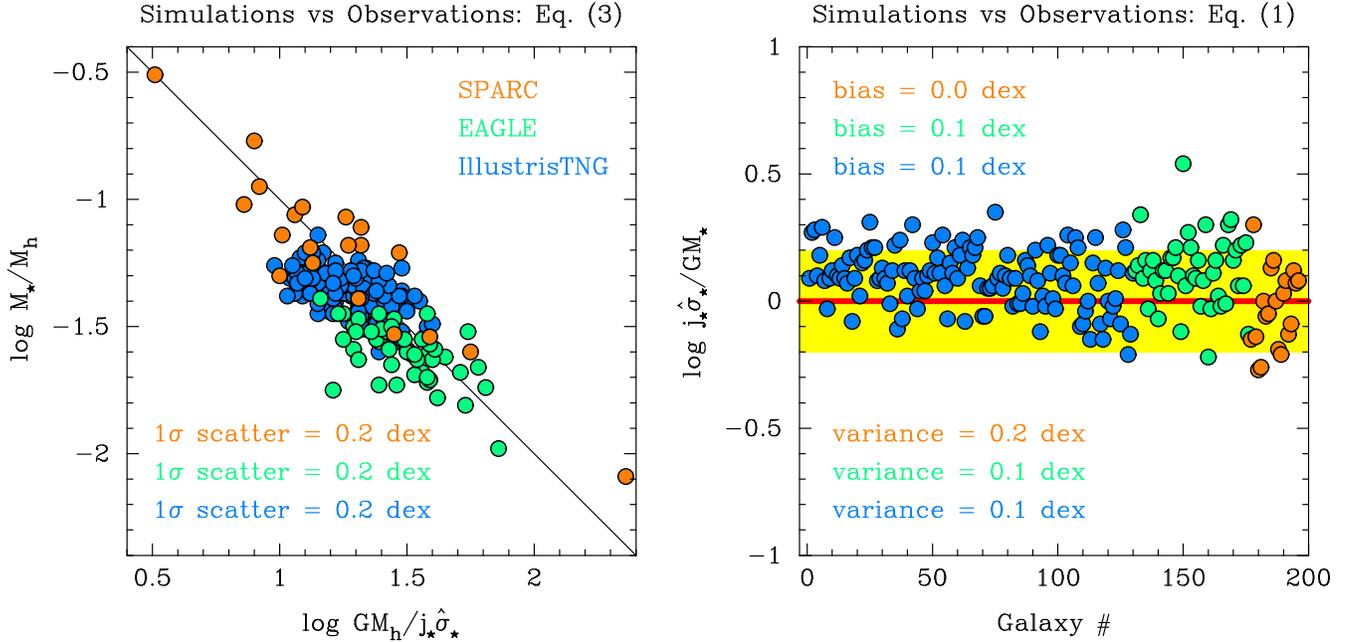}
\caption{Our diagnostics as a critical test for simulations of galaxy
  formation and evolution: the case of central, regularly rotating, massive
  ($M_{\star}>5\times10^{10}\,\mbox{M}_{\odot}$) disc galaxies at redshift
  $z=0$.  The lines in the left and right panels are the predictions based on
  disc gravitational instability.  The shaded region in the right panel is
  the observed $1\sigma$ scatter (0.2 dex for both Eqs 1 and 3), measured
  considering the whole galaxy sample described in Sect.\ 2.  More
  information is given in Sect.\ 4.}
\end{figure*}

We focus on central, regularly rotating, massive
($M_{\star}>5\times10^{10}\,\mbox{M}_{\odot}$) disc galaxies at redshift
$z=0$, and compare the result from the SPARC dataset with the results from
two recent cosmological hydrodynamical simulations: run Ref-L0100N1504 of
EAGLE (Schaye et al.\ 2015), and run TNG100-1 of IllustrisTNG (Pillepich et
al.\ 2018).  We use the three galaxy samples analysed by M20, which were
carefully selected so as to make such a comparison unbiased (see sect.\ 2 of
M20).  In particular, such galaxy samples cover a similar range of stellar
masses ($M_{\star}\approx10^{10.7\mbox{--}11.4}\,\mbox{M}_{\odot}$), hence
they probe the same regime of the stellar mass function.  M20 provided
accurate measurements not only of $M_{\star}$ and $M_{\mathrm{h}}$ but also
of $V_{\mathrm{flat}}$, the circular speed at which the rotation curve
flattens, and $R_{\mathrm{eff}}$, the half-mass radius (see tables A.1 and
A.2 of M20).  This allows us to estimate the stellar specific angular
momentum adopting a commonly used approximation:
$j_{\star}=1.19\,R_{\mathrm{eff}}V_{\mathrm{flat}}$ (e.g., Romanowsky \& Fall
2012).  In reality, more accurate measurements of $j_{\star}$ are available
for the SPARC sample (see Sect.\ 2), but we estimate $j_{\star}$ adopting the
approximation above for consistency with the EAGLE and IllustrisTNG samples.
Finally, we evaluate the velocity dispersion--based quantity
$\hat{\sigma}_{\star}$ using Eq.\ (2) [$i=\star$], as motivated in
Sect.\ 3.2.

Fig.\ 4 provides compelling evidence that simulated and observed galaxies
differ significantly in two complementary respects:
\begin{itemize}
\item The most striking result is the diverse distribution of galaxies
  \emph{along} our scaling relation (see the left panel of Fig.\ 4).
  IllustrisTNG galaxies cluster around
  $M_{\star}/M_{\mathrm{h}}=4.3\times10^{-2}$, while EAGLE galaxies cluster
  around $M_{\star}/M_{\mathrm{h}}=2.5\times10^{-2}$.  This is in sharp
  contrast to SPARC galaxies, which stretch over a range of two orders of
  magnitude and have median $M_{\star}/M_{\mathrm{h}}=6.6\times10^{-2}$.
  Thus such simulations not only underestimate the median of
  $M_{\star}/M_{\mathrm{h}}$, as pointed out by M20, but also dramatically
  misrepresent the large covariance of $M_{\star}/M_{\mathrm{h}}$ and
  $GM_{\mathrm{h}}/j_{\star}\hat{\sigma}_{\star}$.
\item The other result concerns the scatter \emph{across} our scaling
  relation (see again the left panel of Fig.\ 4).  On the one hand, simulated
  galaxies have the same $1\sigma$ scatter as observed ones, which is 0.2 dex
  regardless of whether we consider SPARC galaxies or the whole galaxy sample
  analysed in Sect.\ 3.  On the other hand, simulated galaxies are
  systematically offset from our prediction (the diagonal line), hence their
  `scatter' is partly variance and partly bias.  To illustrate this result
  more eloquently, we make use of the key Eq.\ (1) [$i=\star$] (see now the
  right panel of Fig.\ 4).  IllustrisTNG and EAGLE galaxies show equal
  amounts of variance and bias, whereas SPARC galaxies show twice as much
  variance and no bias at all.  Thus such simulations produce discs that are
  too gravitationally stable and have too little variance in $Q_{\star}$, the
  stellar Toomre parameter.
\end{itemize}

The mismatch in statistical properties between simulations and observations
demonstrates the challenge of reproducing the efficiency of galaxy formation.
This could be due to the inability of current cosmological-volume simulations
to resolve the structure of the ISM, which is crucial for capturing not only
the turbulent nature of star formation but also the subsequent coupling
between feedback and (inter)galactic scales.  Furthermore, resolving the
fragmented nature of the gas affects how AGN fuelling proceeds.  While these
are challenges for present-day models of galaxy formation and evolution, they
will be resolved by future generations of simulations, which aim to cover a
wider dynamical range.  However, the issue of choosing adequate sub-grid
recipes will remain even with improved numerical resolution.  Unless this
problem gets eventually solved by implementations reaching the predictive
level of first principles, sub-grid recipes will still be potentially
important sources of discrepancy between simulated and observed statistical
properties of galaxies, in particular their formation efficiency.

\section{CONCLUSIONS}

Using high-quality data for late-type galaxies of all morphological types
(S0--BCD) from the SPARC and the LITTLE THINGS samples, we have performed a
detailed comparative analysis of three scaling relations: two popular forms
of the SHMR, $M_{\star}/M_{\mathrm{h}}$ vs $M_{\mathrm{h}}$ and
$M_{\star}/M_{\mathrm{h}}$ vs $M_{\star}$, and a scaling relation driven by
disc gravitational instability, $M_{\star}/M_{\mathrm{h}}$ vs
$GM_{\mathrm{h}}/j_{\star}\hat{\sigma}_{\star}$, where $j_{\star}$ is the
stellar specific angular momentum and $\hat{\sigma}_{\star}$ is the stellar
radial velocity dispersion (properly averaged and rescaled).  Our major
conclusions are pointed out below.
\begin{itemize}
\item $M_{\star}/M_{\mathrm{h}}$ vs $M_{\mathrm{h}}$ has a large scatter (0.5
  dex), which hides weak residual trends with stellar mass and gas mass
  fraction.  $M_{\star}$ is a more reliable estimator of
  $M_{\star}/M_{\mathrm{h}}$ than $M_{\mathrm{h}}$: it is unbiased, and it
  also has less scatter (0.3 dex).  This is true regardless of whether we
  consider the whole galaxy sample or the galaxies with most accurate
  measurements of $M_{\star}$ and $M_{\mathrm{h}}$.  Thus
  $M_{\star}/M_{\mathrm{h}}$ vs $M_{\mathrm{h}}$ and
  $M_{\star}/M_{\mathrm{h}}$ vs $M_{\star}$ are not just two alternative
  parametrizations of the same relation, but two significantly different
  relations.
\item $M_{\star}/M_{\mathrm{h}}$ vs
  $GM_{\mathrm{h}}/j_{\star}\hat{\sigma}_{\star}$ has a small scatter (0.2
  dex), which hides a very weak residual trend with gas mass fraction.  But
  this is a second-order effect that vanishes if we consider the galaxies
  with most accurate measurements of $M_{\star}$, $M_{\mathrm{h}}$ and
  $j_{\star}$.  The tightness of such a relation is impressive (see the
  top-left panel of Fig.\ 2), all the more so because the predicted stellar
  mass fraction is not a best-fitting model but results from a parameter-free
  theoretical approach (Romeo 2020).
\item The only difference between $M_{\star}/M_{\mathrm{h}}$ vs
  $GM_{\mathrm{h}}/j_{\star}\hat{\sigma}_{\star}$ and
  $M_{\star}/M_{\mathrm{h}}$ vs $M_{\mathrm{h}}$ is the extra factor
  $G/j_{\star}\hat{\sigma}_{\star}$.  So, although our scaling relation is
  perhaps not a proper SHMR since it cannot predict the halo mass
  explicitely, its tightness tells us that specific angular momentum and
  velocity dispersion may play significant roles in the galaxy-halo
  connection.  In this exploratory work, we have disentangled such roles and
  shown that they are equally important.  In fact, it is the interplay
  between such fundamental galaxy properties that constrains the SHMR so
  tightly in our scenario.
\item Our scaling relation and the key Eq.\ (1) provide a critical test for
  simulations of galaxy formation and evolution.  In this paper, we have
  illustrated how to make use of such diagnostics, and applied them to two
  recent cosmological hydrodynamical simulations: EAGLE and IllustrisTNG.
  Our diagnostics demonstrate that such simulations fail to reproduce several
  statistical properties of the galaxy-halo connection (see Fig.\ 4), which
  ultimately demonstrates how challenging it is to reproduce the efficiency
  of galaxy formation (see Sect.\ 4).
\end{itemize}

Such results have an important astrophysical implication.  Since our scaling
relation originates from the low galaxy-to-galaxy variance of Toomre's (1964)
$Q$ stability parameter, the tightness of such a relation implies that the
process of disc gravitational instability is, on average, remarkably uniform
across disc galaxies of all morphological types: from lenticulars to blue
compact dwarfs.  This is intriguing, as we discuss below.

On the one hand, we know that the stellar Toomre parameter, $Q_{\star}$, is a
reliable disc instability diagnostic in star-dominated galaxies like
lenticulars and spirals (Romeo \& Fathi 2016; Romeo \& Mogotsi 2017; Marchuk
2018; Marchuk \& Sotnikova 2018), provided that the value of $Q_{\star}$ is
properly corrected so as to include disc thickness effects (e.g., Vandervoort
1970; Romeo \& Falstad 2013), and properly interpreted considering the
influence of non-axisymmetric perturbations (e.g., Lau \& Bertin 1978; Griv
\& Gedalin 2012; see sect.\ 5.2 of Romeo \& Fathi 2015 for an updated
discussion).

On the other hand, we know that $Q_{\star}$ is strongly coupled to
$Q_{\mathrm{gas}}$ in gas-rich galaxies (e.g., Lin \& Shu 1966; Romeo \&
Falstad 2013), and that the value of $Q_{\mathrm{gas}}$ is affected by
complex phenomena such as gas turbulence (Romeo et al.\ 2010; Romeo \& Agertz
2014), gas dissipation (Elmegreen 2011), stellar feedback (Agertz et
al.\ 2015), and other non-linear processes (Inoue et al.\ 2016).  The impact
of such phenomena becomes particularly strong in gas-dominated galaxies like
dwarf irregulars and blue compact dwarfs (Renaud 2019), where even the
concept of `disc' is not well defined.

Thus it is really intriguing that all such complexity averages out in a
statistical sense, and that galaxies of such diverse morphology line up along
the same scaling relation driven by disc gravitational instability!  This is
true not only for the SHMR but also for the atomic gas--to--stellar mass
relation (see fig.\ 4 of Romeo 2020), which demonstrates that the cosmic
variance of $Q$ is 0.2 dex, a universal value for both stars and atomic gas.

The astrophysical implication discussed above is a fundamental aspect of
galaxy formation and evolution that deserves to be probed further.  It is not
an easy task to disentangle how much dynamical processes such as star
formation, galaxy minor mergers and satellite accretion have contributed to
the present-day value of $M_{\star}$, hence $M_{\star}/M_{\mathrm{h}}$:
whatever happens during that evolution, disc gravitational instability
regulates $j_{\star}$ and $\hat{\sigma}_{\star}$ for a given $M_{\star}$ so
as to fulfil Eq.\ (1) [$i=\star$],
$j_{\star}(t)\hat{\sigma}_{\star}(t)\approx GM_{\star}(t)$, thus erasing the
memory of those processes.  It is not an easy task, but it can be done by
means of properly designed numerical simulations and/or semi-analytic
modelling, which we challenge the astrophysical community to perform.

\section*{ACKNOWLEDGEMENTS}

We are very grateful to Lorenzo Posti for help with the data, and to Robert
Nau for useful discussions.  We are also very grateful to an anonymous
referee for insightful comments and suggestions, and for encouraging future
work on the topic.

\section*{DATA AVAILABILITY}

The data underlying this article will be shared on reasonable request to the
corresponding author.

\bsp

\label{lastpage}

\end{document}